\documentclass[aps,prb,reprint,superscriptaddress,amsmath,amssymb]{revtex4-2}

\usepackage{etoolbox}% http://ctan.org/pkg/etoolbox
\usepackage{mathrsfs}  
\usepackage{graphicx,amssymb,bm,amsfonts,amsmath,graphics,color,epstopdf}
\usepackage[bookmarks=false,pdfstartview=FitH,colorlinks=true,citecolor=blue,linkcolor=blue]{hyperref}
\usepackage[version=3]{mhchem}

\AtBeginEnvironment{align}{\setcounter{subeqn}{0}}% Reset subequation number at start of align
\newcounter{subeqn} %

\begin{document}

\title{Indirect Excitons and Many-body Interactions in InGaAs Double Quantum Wells}

\author{Christopher L.\ Smallwood}
\email[Email: ]{christopher.smallwood@sjsu.edu}
\affiliation{Department of Physics and Astronomy, San Jos\'e State University, San Jos\'e, CA 95192, USA}
\affiliation{Department of Physics, University of Michigan, Ann Arbor, MI 48109, USA}
\affiliation{JILA, University of Colorado \& National Institute of Standards and Technology, Boulder, CO 80309, USA}
\author{Rachel Owen}
\affiliation{Department of Physics, University of Michigan, Ann Arbor, MI 48109, USA}
\author{Matthew W.\ Day}
\altaffiliation[Present address: ]{Max Planck Institute for the Structure and Dynamics of Matter, Hamburg, Germany}
\affiliation{Department of Physics, University of Michigan, Ann Arbor, MI 48109, USA}
\affiliation{Department of Physics, University of Colorado, Boulder, CO 80309, USA}
\author{Takeshi Suzuki}
\altaffiliation[Present address: ]{Institute for Solid State Physics, University of Tokyo, Kashiwa, Chiba 277-8581 Japan}
\affiliation{Department of Physics, University of Michigan, Ann Arbor, MI 48109, USA}
\affiliation{JILA, University of Colorado \& National Institute of Standards and Technology, Boulder, CO 80309, USA}
\author{Rohan Singh}
\affiliation{Department of Physics, University of Michigan, Ann Arbor, MI 48109, USA}
\affiliation{JILA, University of Colorado \& National Institute of Standards and Technology, Boulder, CO 80309, USA}
\affiliation{Department of Physics, University of Colorado, Boulder, CO 80309, USA}
\affiliation{Department of Physics, Indian Institute of Science Education and Research Bhopal, Bhopal 462066, India}
\author{Travis M.\ Autry}
\affiliation{JILA, University of Colorado \& National Institute of Standards and Technology, Boulder, CO 80309, USA}
\affiliation{Department of Physics, University of Colorado, Boulder, CO 80309, USA}
\author{Smriti Bhalerao}
\affiliation{Department of Physics and Astronomy, San Jos\'e State University, San Jos\'e, CA 95192, USA}
\author{Fauzia Jabeen}
\affiliation{Univ Rennes, INSA Rennes, CNRS, Institut FOTON-UMR 6082, F-35000, Rennes, France}
\affiliation{Laboratory of Quantum Optoelectronics, \'Ecole Polytechnique F\'ed\'erale de Lausanne (EPFL), CH-1015 Lausanne, Switzerland}
\author{Steven T.\ Cundiff}
\email[Email: ]{cundiff@umich.edu}
\affiliation{Department of Physics, University of Michigan, Ann Arbor, MI 48109, USA}
\affiliation{JILA, University of Colorado \& National Institute of Standards and Technology, Boulder, CO 80309, USA}
\affiliation{Department of Physics, University of Colorado, Boulder, CO 80309, USA}
\date {\today}

\begin{abstract}
Spatially indirect excitons in semiconductor quantum wells are relevant to basic research and device applications because they exhibit enhanced tunability, delocalized wave functions, and potentially longer lifetimes relative to direct excitons. Here we investigate the properties of indirect excitons and their coupling interactions with direct excitons in asymmetric \ce{InGaAs} double quantum wells using optical multidimensional coherent spectroscopy and photoluminescence excitation spectroscopy. Analyses of the spectra confirm a strong influence of many-body effects, and reveal that excited-state zero-quantum coherences between direct and indirect excitons in the quantum wells dephase faster than the much higher-energy single-quantum coherences between excitonic excited states and ground states. The results also suggest an important energy-dependent role of continuum states in mediating system dynamics, and they indicate that dephasing mechanisms are associated with uncorrelated or anticorrelated energy-level fluctuations. 
\end{abstract}

\maketitle

\section{Introduction}

Quantum confinement is a phenomenon where the dimensions of an object become small enough to measurably impact intrinsic energy levels. The effect in turn has exerted tremendous influence over the course of recent technological progress. One-dimensional confinement, as realized through quantum wells, has facilitated the manufacture of efficient light-emitting diodes and lasers \cite{Faist1994,Nakamura1995,Nakamura1995a}. Three-dimensional confinement, as realized through quantum dots, has transformed biomedical imaging \cite{Zrazhevskiy2010} and illuminated display technologies \cite{Kim2024}. It has also led to promising developments in quantum information technology \cite{Arakawa2020,Heindel2023}.

Alongside industrial applications, the ability to generate and manipulate semiconductor quantum confinement has led to important advances in fundamental science, aided by the enhanced level of control that confinement effects facilitate. For example, the fabrication and manipulation of coupled quantum wells with tailored interwell barrier thickness allows one to investigate tunneling and hybridization. Tunneling between quantum wells is relevant to light-matter interaction processes in semiconductors \cite{Tsuchiya1987,Fox1991} and energy transfer processes in photosynthetic light harvesting complexes \cite{Engel2007,Cao2020}. Hybridization effects are relevant to the generation and manipulation of spatially indirect excitons \cite{Butov1995,Bayer1996,Voros2005}, which straddle pairs of quantum wells with electron components primarily residing in one well and hole components primarily residing in the other. Populations of indirect excitons have been proposed to realize Bose-Einstein condensates in III-V semiconductors \cite{Combescot2017} and have been demonstrated to realize long-lived resonance features in transition metal dichalcogenide heterostructures \cite{Fang2014,Chiu2014,Rivera2015,Jiang2021}.  

Research efforts aiming to better understand the features of direct and indirect excitons as they inhabit and tunnel between quantum well structures are ongoing. They stand to benefit from refined sample growth methods \cite{Herman,OuelletPlamandon2015,Dong2024} and sophisticated optical spectroscopy techniques \cite{Li,Nardin2014,Tollerud2016,Tollerud2017}. To that end, in this article we report measurements using optical multidimensional coherent spectroscopy (MDCS) and photoluminescence excitation (PLE) spectroscopy to probe the coherent properties of both direct and indirect excitons in asymmetric \ce{InGaAs} double quantum wells. We have investigated a collection of three samples where the barrier width between a 9 nm and 10 nm quantum well has been grown at 5 nm, 10 nm, and 30 nm, and we have characterized the emergence of optically accessible indirect-exciton transitions in the system as the barrier thickness decreases. Among the most prominent findings of the study is an observation that zero-quantum coherences dephase more quickly than single-quantum coherences, and an observation that excitation-induced dephasing (EID) effects tend to increase with increasing exciton energy relative to excitation-induced shift (EIS) effects for direct and indirect excitons alike. Results indicate that decoherence mechanisms are likely rooted in uncorrelated or anticorrelated energy-level fluctuations with continuum states playing an increasingly important role in this process at higher energies. In broader context, the results may be relevant to the physics of manufactured devices including quantum cascade lasers, and to transfer efficiency between energy levels in a variety of naturally occurring quantum confined systems.

\section{Sample Characteristics and Peak assignments}
\label{sec:ple}

\begin{figure}[tb]\centering\includegraphics[width=3.375in]{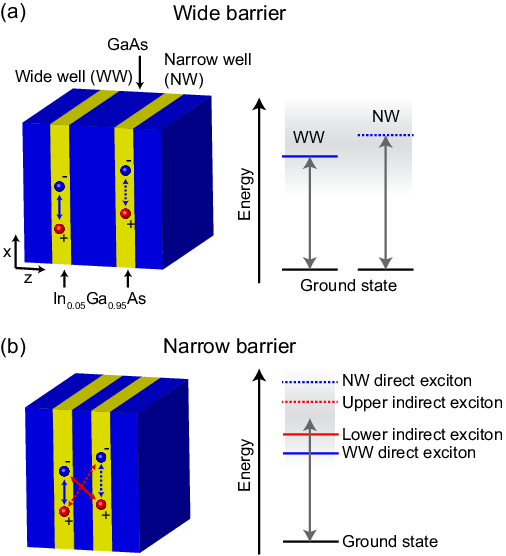}
\caption{Low-energy exciton states in asymmetric \ce{InGaAs/GaAs} double quantum wells. The samples consist of two \ce{In_{0.05}Ga_{0.95}As} quantum wells of respective thickness 10 nm [wide well (WW)] and 9 nm [narrow well (NW)], separated by a \ce{GaAs} interwell barrier.
{\bf(a)} For a wider barrier, the elementary excitations are direct excitons, in which the electron and hole constituents reside in the same quantum well.
{\bf(b)} As the barrier narrows, indirect excitons---in which the electron and hole constituents reside in different quantum wells---become increasingly visible.}
\label{cartoon}
\end{figure}

InGaAs double quantum well samples were fabricated using molecular beam epitaxy (MBE) by growing layers of \ce{InGaAs} with 5\% indium concentration between layers of pure \ce{GaAs} to create structures as illustrated in Fig.~\ref{cartoon}. The samples constitute useful testbeds for exploring exciton physics because confinement effects associated with the well thickness (10 nm for the ``wide well" and 9 nm for the ``narrow well") move the optical resonances associated with the two different wells into spectrally separable energy ranges. In addition, the barrier thickness between quantum wells is relevant to indirect exciton generation efficiency. In the wide-barrier limit [Fig.~\ref{cartoon}(a)], a vanishing spatial overlap between electron and hole wave functions in separate wells prohibits indirect exciton generation by light and---although indirect excitons could in principle still exist---the only optically accessible bound states are direct excitons where the electron and hole components primarily reside within the same quantum well. As the barrier width narrows, indirect transitions become increasingly prominent. The result is a crossover from two optically accessible transitions into four transitions, as depicted by the energy level diagrams on the righthand sides of Figs.\ \ref{cartoon}(a) and \ref{cartoon}(b). Transitions here involve the heavy-hole band of \ce{InGaAs} only (the transition involving the light-hole band is pushed outside of the observable spectral window in accordance with the large strain applied to the quantum wells, due in turn to the lattice mismatch between \ce{GaAs} and \ce{InGaAs}). All measurements were conducted at 10 K, thereby minimizing the impact of thermal broadening.

%%%%%%%%%%%%%%%%%%%%%%%%%%%%%%%%%
%Fig1
%%%%%%%%%%%%%%%%%%%%%%%%%%%%%%%%%
%Figure source: Local eps. Fig1.ai looks to be okay too back on 11/29/2022.

\begin{figure}[tb]\centering\includegraphics[width=3.375in]{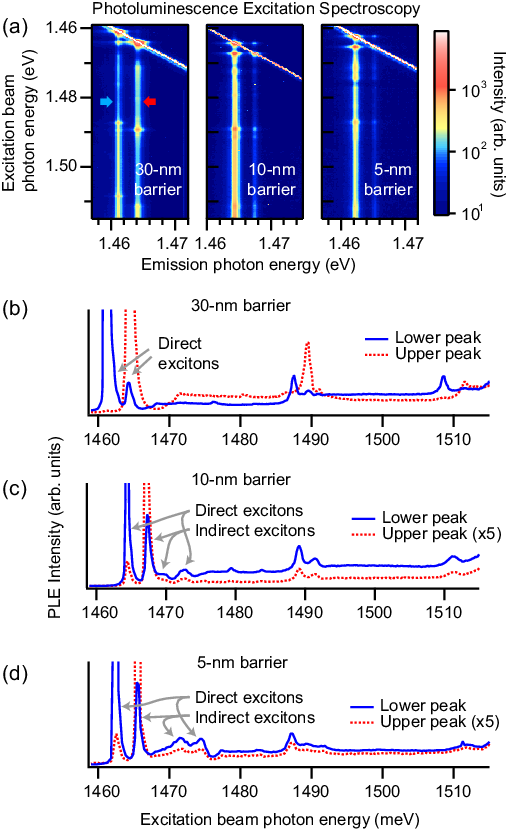}
\caption{Photoluminescence excitation spectroscopy (PLE) measurements of asymmetric double-quantum wells.
{\bf(a)} Two-dimensional PLE plots of photoluminescence intensity vs.\ emission frequency and excitation frequency. The vertical emission streaks are the two lowest-energy exciton transitions in each of the samples. The diagonal streak at excitation frequency = emission frequency originates from laser scatter.
{\bf(b)} Photoluminescence vs.\ excitation frequency at fixed emission frequency, corresponding to the two lowest energy emission peaks in the 30-nm-barrier sample.  
{\bf(c)} Same as (b) for the 10-nm-barrier sample.
{\bf(d)} Same as (b) for the 5-nm-barrier sample.
}
\label{ple}
\end{figure}

\begin{figure*}[tb]\centering\includegraphics[width=6.5in]{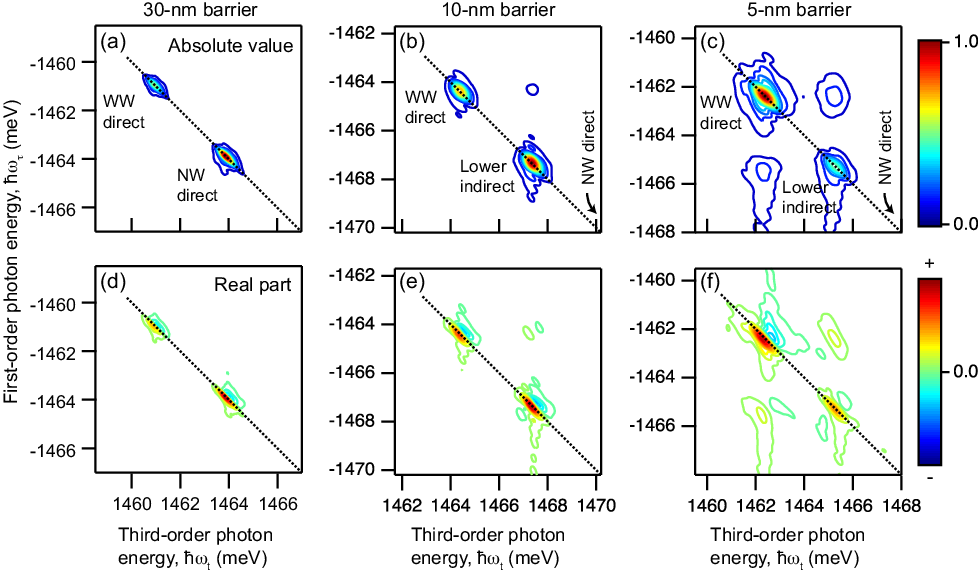}
\caption{\label{phaseresolved}Phased-resolved single-quantum rephasing MDCS plots where the laser bandwidth has been tuned to accentuate the two lowest-energy excitons. {\bf (a)}--{\bf (c)} Absolute value spectra. {\bf (d)}--{\bf (f)} Real parts of the spectra.
}
\end{figure*}

To initially characterize the emergence of indirect transitions in these systems and to verify peak assignments, we conducted PLE measurements, acquired by scanning the frequency of a continuous-wave (CW) Ti:sapphire laser across the optical transitions (see the gray double arrows on the righthand side of Fig.~\ref{cartoon}). Figure \ref{ple} shows the results. As shown in Fig.~\ref{ple}(a), two-dimensional plots as a simultaneous function of the frequency of excitation light and the frequency of emitted photoluminescence (PL) reveal vertical streaks of intensity corresponding to emission from the system's lowest-energy excitons. Data sets are plotted with the vertical axis increasing in value downward to highlight parallels with the data presented in Figs.~\ref{offres} and \ref{phaseresolved} later on. Among the advantages of PLE spectroscopy is that exciton transitions at energies well above the lowest-energy states can be characterized by examining the changing intensity of an emission peak as a function of varied excitation frequency; i.e., by extracting vertical lineouts of the data in Fig.~\ref{ple}(a) at fixed values of emission photon energy as illustrated in Figs.~\ref{ple}(b)--\ref{ple}(d). The measurement shares characteristics with linear absorption, but in many ways is better because the emission-frequency filtering facilitates the isolation of phenomena specific to the wider well or narrower well in cases where the two wells are uncoupled.

%%%%%%%%%%%%%%%%%%%%%%%%%%%%%%%%%
%Fig2
%%%%%%%%%%%%%%%%%%%%%%%%%%%%%%%%%
%Figure source: Local eps.
%Data source: 2018-02-09 PLE briefer.pxp, saved onto Chris's desktop.
%New data source: 2025-01-03 PLE briefer.pxp, saved onto Chris's desktop.

An effect of exactly this sort occurs in Fig.~\ref{ple}(b), where the excitation spectra derived from the 1461-meV
%Exact fit: 1461.1 meV 
and 1464-meV
%Exact fit: 1464.1 meV
emission peaks [see the blue and red arrows in Fig.~\ref{ple}(a)] generate very different profiles. The low-energy emission peak spectrum [solid blue trace, Fig.~\ref{ple}(b)] corresponds primarily to wide-well transitions, where the exciton corresponding to the $n=1$ interband quantum-well transition dominates the low-energy spectrum with a peak exceeding the scale of the vertical axis at excitation photon energy 1461.1 meV (i.e., resonant excitation of the wide-well exciton). Above this, near 1464 meV, there is a smaller peak that likely originates from weak dipole-dipole energy transfer processes between the wide well and narrow well \cite{Tomita1992,Tomita1996,Borri1997}. Above this is a spectral shoulder at 1468.3 meV, which corresponds to the onset of wide-well continuum states associated with the $n=1$ transition and establishes an excitonic binding energy of 7.2 meV\@. Above this yet further, several other higher-energy exciton resonances overlap against the $n=1$ state continuum. For example, the cusp at 1476.3 meV corresponds to the parity-forbidden transition coupling the $n=1$ electron state to an $n=2$ hole state. The other higher-energy states correspond to indirect barrier excitons, which have coherent properties that have been characterized in more detail in a related sample by the authors of Ref.~\cite{Tollerud2016}. The higher-energy emission peak spectrum [dotted red trace, Fig.~\ref{ple}(b)] corresponds primarily to narrow-well transitions, and exhibits a corresponding set of narrow-well peak assignments with a lowest-energy exciton binding energy of 7.4 meV\@.

Figures \ref{ple}(c) and \ref{ple}(d) show vertical lineouts of the PLE spectra from the 10-nm-barrier and 5-nm-barrier samples. The spectra demonstrate emergent interwell coupling features. Whereas the exciton peaks that are visible in Fig.\ \ref{ple}(b) appear at quite different energies between the ``lower peak" blue trace and the ``higher peak" red trace, these peak profiles become much more similar in Figs.\ \ref{ple}(c) and \ref{ple}(d). Moreover, the red traces in Figs.\ \ref{ple}(c) and \ref{ple}(d) are significantly attenuated relative to their blue trace counterparts, which is an indicator that decreasing the quantum-well barrier thickness opens up interaction channels that are not accessible to excitons in the wide-barrier sample.

A particularly noteworthy feature of Figs.\ \ref{ple}(c) and \ref{ple}(d) is the splitting of Fig.\ \ref{ple}(b)'s two lowest-energy exciton peaks into what become four clearly defined low-energy exciton peaks in Fig.\ \ref{ple}(d) at 1462.6 meV, 1465.6 meV, 1471.7 meV, and 1474.5 meV\@. The lowest and highest of these resonances correspond to direct excitons in the 5-nm-barrier sample, while the middle two resonances are signatures of indirect exciton emergence. The energy separation of 3.0 meV between the lowest-energy direct exciton and lowest-energy indirect exciton is well below the expected exciton binding energies (7.2--7.4 meV, based on the 30-nm-barrier sample measurements), allowing us to examine interactions between these resonant features in a fairly direct manner. Beyond the exploration of narrow barrier widths, this constitutes one of the main advantages of the present study over previous MDCS work on other double-quantum well samples \cite{Nardin2014,Tollerud2016,Tollerud2017}, where the differences between quantum well thicknesses were bigger and the minimal exciton energy separations were on par with exciton binding energies.

\section{Direct and Indirect Exciton Coherent Coupling}
\label{sec:offres}

Having established peak assignments for the direct and indirect exciton resonant frequencies, we proceed to a discussion of coherent coupling effects. To access these, we conducted wave-vector-selection-based multidimensional coherent spectroscopy (MDCS) measurements \cite{Li} using a custom-built spectrometer \cite{Bristow2009}. For the implementation of MDCS described in this article, three successive laser excitation pulses were directed upon the sample to create a coherent four-wave mixing emission signal, which was then amplified and phase-resolved through heterodyne detection involving a local oscillator pulse that was routed around the sample to eliminate undesirable sample absorption effects. Conventions are such that the time following the first-order excitation pulse is labeled $\tau$ (with conjugate variable $\omega_\tau$), the time following the second-order excitation pulse is labeled $T$, and the time following the third-order excitation pulse is labeled $t$ (with conjugate variable $\omega_t$). We calibrated the global phase of our measurements using auxiliary pump-probe measurements. MDCS excitation pulses were near-infrared laser pulses generated by a mode-locked Ti:sapphire oscillator with a repetition rate of 76 MHz and a bandwidth of 18 meV\@. We collected spectra from rephasing pulse sequence emission signals (i.e., ${\bf k}_{sig} = - {\bf k}_1 + {\bf k}_2 + {\bf k}_3$, where the subscript indicates the time ordering of each of three different excitation pulses).

\begin{figure}[tb]\centering\includegraphics[width=3.375in]{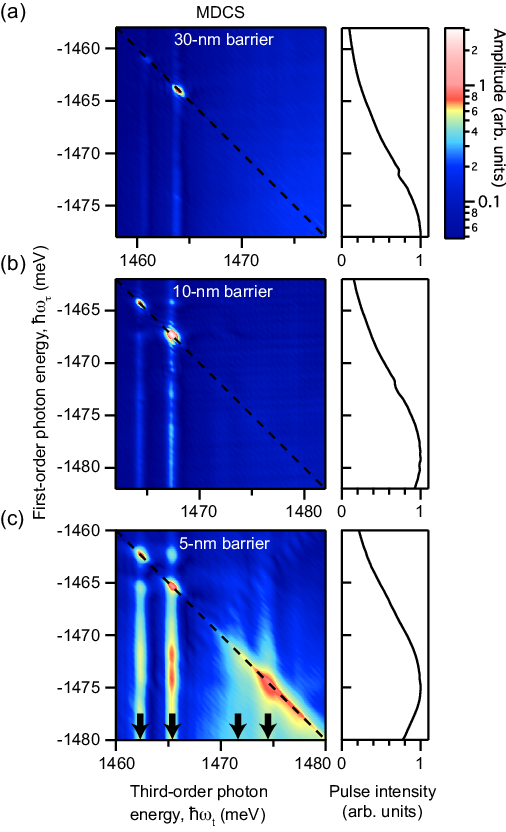}
\caption{Multidimensional coherent spectroscopy (MDCS) measurements of asymmetric double-quantum wells. The two-dimensional color plots correspond to rephasing single-quantum spectra, acquired with  excitation pulses tuned 12--17 meV above the lowest-energy resonance peaks in order to highlight higher-energy exciton states. The graphs on the righthand side of each of the panels illustrate the excitation laser power spectra.
{\bf(a)} 30-nm barrier sample.
{\bf(b)} 10-nm barrier sample.
{\bf(c)} 5-nm barrier sample.
}
\label{offres}
\end{figure}

Figure \ref{offres} shows single-quantum MDCS rephasing measurements at $T = 200$ fs resulting from an experiment where we intentionally tuned the excitation laser above the lowest-energy exciton states by about 15 meV to highlight the full range of direct and indirect exciton states. As illustrated by the peaks along the diagonal line $(\hbar \omega_t = -\hbar \omega_\tau)$ at 1461 meV and 1464 meV in Fig.\ \ref{offres}(a), the MDCS measurements are in agreement with the low-energy PLE peaks in Fig.\ \ref{ple}. Figures \ref{offres}(b) and \ref{offres}(c) show MDCS measurements of 10-nm-barrier and 5-nm-barrier samples, respectively, and---as was the case with the PLE measurements---the separation between low- and high-energy direct exciton peaks widens as the barrier thickness between wells decreases. Indirect exciton peaks begin to emerge in the photon energies between.

%%%%%%%%%%%%%%%%%%%%%%%%%%%%%%%%%
%Fig3
%%%%%%%%%%%%%%%%%%%%%%%%%%%%%%%%%
%Figure source: Local eps. An earlier version is Fig3_was_OffRes.ai
%Data source: 2018-03-05 OffResonant.pxp, drawing upon data collected 2015-08-17, 2015-08-18, 2015-08-24.

More striking than the diagonal peak features is the emergence of cross-peaks---visible most clearly for the 5-nm-barrier sample in Fig.~\ref{offres}(c)---which appear away from the $\hbar \omega_t = -\hbar \omega_\tau$ diagonal line, and which indicate coherent coupling between exciton states in quantum wells with narrow barriers. Cross-peaks like this have been observed before in Ref.~\cite{Nardin2014}, where the authors concluded that a coupling effect arises from many-body interactions. However, the observation here is in the midst of a much more intertwined mixing of wide-well and narrow-well exciton states than had been possible to observe in Ref.~\cite{Nardin2014}, making the present study distinct. Figure \ref{offres}(c) shows, for example, a significant degree of coupling between all four of the lowest-energy exciton states as evidenced by the cross peaks connecting each of the four diagonal resonances at photon energies 1462.4 meV, 1465.4 meV, 1471.9 meV, and 1474.5 meV [see the black arrows across the bottom axis of Fig.~\ref{offres}(c); resonance photon energies in this plot are consistent with those extracted from PLE peak locations, with an experimental uncertainty for both sets of measurements at about 0.2 meV\@.] In addition to coupling between direct and indirect excitons, coupling between exciton states and quantum-well continuum states is also evident as illustrated by the vertical streaks accompanying these peaks \cite{Borca2005} at third-order energies $\hbar \omega_t = 1462.4$ meV and 1465.4 meV\@.

%%%%%%%%%%%%%%%%%%%%%%%%%%%%%%%%%
%Fig4
%%%%%%%%%%%%%%%%%%%%%%%%%%%%%%%%%
%Figure source: Local eps. An earlier version is Fig4_was_TheoryFig2.ai
%Data source: 2018-03-07 Transition Energies.pxp. Note that the intensity scaling of panel (d) was rendered after the fact in Illustrator.
\begin{figure}[tb]\centering\includegraphics[width=3.375in]{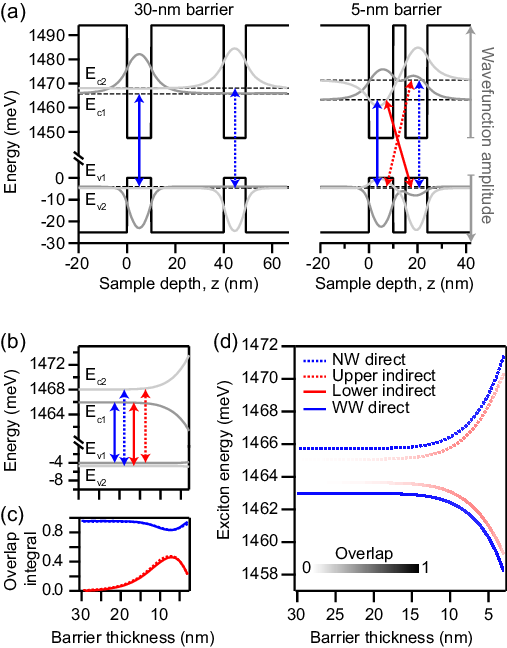}
\caption{\label{theory} Modeling the emergence of indirect excitons. 
{\bf(a)} Solutions for the energies and wave functions of the two highest-lying valence band states ($E_{v1}$ and $E_{v2}$) and lowest-lying conduction band states ($E_{c1}$ and $E_{c2}$) in the double quantum well system, for a 30-nm barrier (left) and 5-nm barrier (right).
{\bf(b)} Energies of the eigenfunctions from (a), vs.\ barrier thickness. The configuration of states leads to the possibility of two direct transitions (blue double arrows) and two indirect transitions (red double arrows).
{\bf(c)} Overlap integrals for the transitions illustrated in (a) and (b).
{\bf(d)} Combined plot of excitonic optical transition energies and overlap integrals. Vertical position corresponds to exciton energy (i.e., the energy difference between the states depicted in (a), subtracting an additional 7 meV for the exciton binding energy). Color saturation corresponds to the overlap integral magnitude. 
}
\end{figure}

The detailed coupling effects connected to the intensities of the cross-peaks in Fig.~\ref{offres} require a discussion of many-body interactions that we shall postpone until Sections \ref{sec:manybody} and \ref{sec:zeroq}, but many of the salient features from Figs.\ \ref{ple} and \ref{offres} regarding the locations and intensities of diagonal peaks can be understood within the context of a simple quantum mechanical treatment of particles restricted in one dimension by finite square well potentials. Figure \ref{theory} shows an illustration of this treatment, and provides an intuitive understanding of both why the energy separation between direct excitons should be expected to widen as interwell barrier thickness decreases, and why the indirect exciton signatures between the two direct exciton signatures should be expected to simultaneously become more prominent. 

As shown in Fig.\ \ref{theory}(a), we can model the system as two sets of double square well potentials (one for the electron in the conduction band and one for the hole in the valence band). The left wells have a thickness of 10 nm, the right wells have a thickness of 9 nm, and the barrier thickness between wells is set as an adjustable parameter. The depth of the wells is different in the case of the electron and hole states, with a band offset of about 0.65 determined by comparing to the PLE data displayed in Fig.~\ref{ple}(b) in a manner similar to that reported in Ref.~\cite{TollerudThesis}. The InGaAs bandgap as determined from this comparison is estimated at 1.447 eV\@. Electron and hole effective mass terms in InGaAs and the GaAs bulk material bandgap at low temperature are drawn from the literature \cite{Ji1987,Vurgaftman2001,Tollerud2016,Blakemore1982}. They correspond to $m_e^*=0.065 \, m_0$ and $m_h^*=0.505 \, m_0$ (where $m_0$ is the electron mass in vacuum) and $E_{g,\textrm{GaAs}} = 1.519$ eV\@.

Having established these parameters, quantum confinement effects along the $z$ direction can be incorporated by means of separation-of-variables techniques \cite{Fox}. The energy shifts of free electron and hole states and $z$-direction wave function profiles can be determined by solving the one-dimensional time-independent Schr\"odinger equation
\begin{equation}
-\frac{\hbar^2}{2m^*} \frac{\partial^2 \psi_n}{\partial z^2} + V(z)\psi_n = E_n \psi_n.
\end{equation}
Using numerical methods of enforcing consistent boundary conditions at each of the well edges, we obtained
the two lowest-energy solutions to this equation for both electron states and hole states. These are plotted on the left and right side of Fig.~\ref{theory}(a) and associated with the energies $E_{c1}$ and $E_{c2}$ (for the energies of the first and second conduction band eigenstates) and $E_{v1}$ and $E_{v2}$ (for the energies of the first and second valence band eigenstates). 

Because electrons in the system have a much smaller effective mass than holes, one can immediately see from the analysis that a primary effect of decreasing barrier thickness in the sample is on the energies of the electron states. As shown in Fig.~\ref{theory}(b), the hole energies $E_{v1}$ and $E_{v2}$ at the top of the valence band are almost constant functions of barrier thickness. By contrast, the electron energies $E_{c1}$ and $E_{c2}$ at the bottom of the conduction band veer sharply away from each other at barrier thicknesses less than 10 nm.

Another prominent feature of Fig.~\ref{theory}(a) is that the wave functions become increasingly delocalized in the $z$ direction as a function of decreasing barrier thickness. The effect is illustrated by the gray curves in the top right portion of Fig.~\ref{theory}(a) in comparison to their counterparts in the top left portion of Fig.~\ref{theory}(a). The delocalization effect has an impact on the overlap integral 
\begin{equation}
\int_{-\infty}^\infty \psi^*_{vn}(z)\psi_{cm}(z) dz,
\end{equation}
and the squared amplitude of this integral is in turn proportional to the transition probability for exciting a given exciton \cite{Fox}. Overlap integral values are plotted as a function of barrier thickness in Fig.~\ref{theory}(c).

%%%%%%%%%%%%%%%%%%%%%%%%%%%%%%%%%
%Fig5
%%%%%%%%%%%%%%%%%%%%%%%%%%%%%%%%%
%Data source: 2023-03-29 Phased Resonant.pxp, drawing from 2015-07-10, 2015-07-13, 2015-07-14, and 2018-04-05.

Figure \ref{theory}(d) shows a combined exciton transition probability plot as extracted from the information displayed in Figs. \ref{theory}(b) and \ref{theory}(c). In order to approximate the attractive electron-hole binding energy, the transitions in the panel have been uniformly reduced by a constant 7 meV\@. As the figure shows, only two transitions---associated with direct exciton transitions---are optically accessible in the case of the 30-nm barrier sample. As the barrier thickness decreases, the overlap integrals for the indirect transitions increase. Indirect transitions become visible in addition to the direct transitions. The energies of these four transitions diverge into a wider range of energies in a manner nicely explaining the data presented in Figs.~\ref{ple} and \ref{offres}. 

Alongside its strengths, the simulation has a few notable limitations. The simulated energy separation between the two lowest-energy excitons at a barrier separation of 5 nm is 0.7 meV, whereas the experimental separation between these excitons is 3.0 meV\@. Absolute values of experimental exciton energies are also somewhat variable among the 30-nm, 10-nm, and 5-nm sample in ways that are not captured by the simulation. More sophisticated simulations involving variable exciton binding energies and strain effects may help resolve these discrepancies. Strain, for example, is an intrinsic feature resulting from the lattice mismatch between InGaAs and GaAs, and may shift the overall spectrum of excitons up or down, leading to coincidences where the lowest-energy exciton appears not to shift in energy while the higher-energy excitons appear to move by large amounts. These kinds of effects may occur randomly in the system from one growth operation to the next, or as a result of an inhomogeneous distribution of indium atoms as the MBE process proceeds with more (or less) indium being deposited as the quantum well is beginning to be grown in comparison to where a multiple-nanometer fraction of the quantum well has already been accumulated. The approximation here of uniform 7-meV binding energies could also benefit from being refined in response to effects resulting from both strain and varied barrier thickness. It is known, for example, that the binding energies of excitons in GaAs/AlGaAs quantum wells depend on quantum well thickness in nontrivial ways \cite{Greene1984}. The coupled quantum wells here studied may resemble single wells of extended thickness in the case of the narrowest barriers, leading to similar dependencies. Barrier thickness between semiconductor quantum wells has also been shown to play an important role in the dynamics of multimode polaritons \cite{OuelletPlamandon2015}.

\section{Many-body effects}
\label{sec:manybody}

Among the advantages of using a technique like MDCS to examine excitons in semiconductors is the ability of the technique not just to identify peak energies, but to elucidate many-body interaction effects. To assist in achieving this goal, we show in Fig.~\ref{phaseresolved} a series of phase-resolved 2D rephasing spectra in which the frequency of the excitation laser was tuned into resonance with the two lowest-energy excitons. Figure \ref{phaseresolved}(a) shows the absolute value of a single-quantum rephasing spectrum corresponding to the 30-nm-barrier sample, where wide-well and narrow-well direct exciton resonances are clearly distinguishable and [as noted in the discussion of Fig.~\ref{offres}(a)] uncoupled from each other. 

Figures \ref{phaseresolved}(b) and \ref{phaseresolved}(c) illustrate indirect exciton emergence as the barrier width narrows. Note that although peak energies separated by 3 meV still appear in Figs.~\ref{phaseresolved}(b) and \ref{phaseresolved}(c) as they did in Fig.~\ref{phaseresolved}(a), the higher-energy peak in these panels is \textit{not} the narrow-well direct exciton as it was in Fig.~\ref{phaseresolved}(a), but rather the lower-energy indirect exciton (the narrow-well direct exciton gets pushed outside the observation window to higher energies; refer back to Figs.~\ref{ple} and \ref{offres}). At the same time, coupling effects between this lower-energy indirect exciton and the wide-well direct exciton become prominent, as evidenced by cross-peak appearance. Cross-peaks emerge symmetrically in these samples, appearing with nearly equal magnitude both below and above the $\hbar\omega_t = -\hbar\omega_\tau$ diagonal line in Fig.~\ref{phaseresolved}(c). 

\begin{figure}[tb]\centering\includegraphics[width=3.375in]{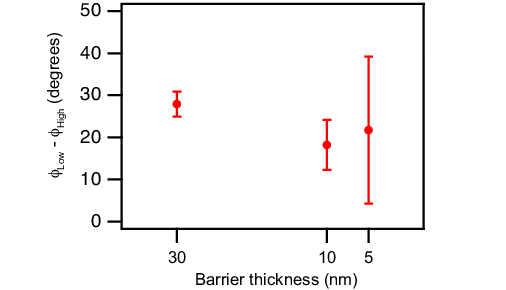}
\caption{\label{phaseresolved2}Phase shift difference between the lowest-energy diagonal resonance and the higher-energy diagonal resonance above it in Fig.~\ref{phaseresolved}, plotted vs.\ barrier thickness.
}
\end{figure}

Figures \ref{phaseresolved}(d)--\ref{phaseresolved}(f) show the real parts of the spectra depicted in Figs.~\ref{phaseresolved}(a)--\ref{phaseresolved}(c) and reveal, similarly to the findings of Ref.~\cite{Nardin2014}, that many-body effects play an important role in generating both diagonal spectral features and cross-peaks. The phase of diagonal spectral peak maxima (for direct and indirect exciton resonances alike) is always positive. This is different from the absent phase shift that would have been predicted in a single-particle model. The shift is evidenced by the fact that peaks in the real parts of spectra come to maximal values below the $\hbar\omega_t = -\hbar\omega_\tau$ diagonal lines and are accompanied by asymmetrically placed negative lobes above the $\hbar\omega_t = -\hbar\omega_\tau$ diagonal lines. The fact that a phase shift exists at all is strong evidence of the role of many-body effects in these materials \cite{Li,Li2006}. The phase shift hovers around 45 degrees in absolute value and is measurement-dependent, likely varying with local strain and also slight variations in sample temperature from one measurement to the next. On a fairly consistent basis, however, the shift of higher-energy peaks tends to be smaller than that of the low-energy peak in the samples measured as illustrated in Fig.~\ref{phaseresolved2}, indicating an enhanced level of EID relative to EIS for higher-energy excitons. (Details of the effect are discussed below in Appendix \ref{sec:phaseshifts}.)

In turn, the increasing prominence of EID in comparison to EIS in higher-energy excitations as opposed to lower-energy excitations indicates an important role played by continuum states in influencing many-body effects, Even though continuum states likely do not directly overlap the higher-energy exciton states, the energy separation is smaller than for the lowest-energy exciton states, leading to an enhancement of interaction channels. Higher-lying Rydberg states of the excitons also likely play a role. These coupling effects are experimentally manifest in the emergence of a vertical streak in the lower-left cross-peak in Fig.~\ref{phaseresolved}(c) [see also the complimentary image of the same cross-peak in Fig.~\ref{offres}(c)], beginning only 3 meV above the wide-well direct exciton energy in spite of the fact that the binding energy of this exciton is expected to be about twice as big as this [refer back to Fig.~\ref{ple}(b)].

%%%%%%%%%%%%%%%%%%%%%%%%%%%%%%%%%
%Fig6
%%%%%%%%%%%%%%%%%%%%%%%%%%%%%%%%%
%Data source: 2023-03-29 Phased Resonant.pxp, drawing from 2015-07-10, 2015-07-13, 2015-07-14, and 2018-04-05.

Evidence for many-body effects is additionally present in the dispersive (i.e., phase-shifted) nature of the cross-peaks displayed in Figs.~\ref{phaseresolved}(f): the real part of the lower-left cross-peak, in particular, features both positive and negative contributions visible within the peak. Such features do not appear in cross-peaks generated by phase-space filling effects. A dispersive cross-peak feature akin to the one illustrated by Fig.~\ref{phaseresolved}(f) was also observed in previous measurements on the related samples measured in Ref.~\cite{Nardin2014}. However, the phases of diagonal peaks as measured by the previous study and the present one are different: the previous study measured primarily absorptive diagonal peaks, whereas the peak phases of the present study are more dispersive. Implications of this difference will be discussed later on in Section \ref{sec:conclusions}. 

\section{Dephasing rates}
\label{sec:zeroq}

In addition to allowing an ability to examine many-body effects by means of phase-resolved spectral signatures, the MDCS technique allows the ability to pull out different kinds of characteristic timescales. Among the more interesting of these in the case of coupled quantum wells is a comparison between the dephasing rates associated with diagonal single-quantum coherences (which can be obtained by measuring the homogeneous linewidths of the data presented in Fig.~\ref{phaseresolved}) and the dephasing rates associated with coherences that can be generated between nearly degenerate excited states. These latter coherences are termed ``zero-quantum" coherences because the states to which they pertain are not connected by a quantum of the electromagnetic field (a photon). They can be measured by collecting MDCS data as a function of varied second-order interaction time $T$.

%%%%%%%%%%%%%%%%%%%%%%%%%%%%%%%%%
%Fig7
%%%%%%%%%%%%%%%%%%%%%%%%%%%%%%%%%
%Data source: 2024-07-25 DQW 3D data5.pxp, saved onto Chris's hard drive. Drawing from older data sets collected in Boulder.
\begin{figure}[tb]\centering\includegraphics[width=3.375in]{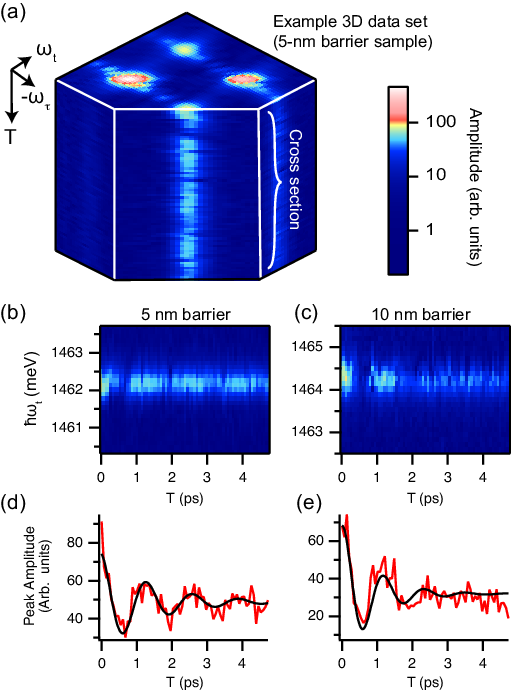}
\caption{Evolution of spectral properties as a function of varied second-order waiting time $T$. \textbf{(a)} Example data set. \textbf{(b)} and \textbf{(c)} Cross sections were extracted by slicing through the lower-left cross-peak in the 5-nm-barrier and 10-nm-barrier data sets. \textbf{(d)} and \textbf{(e)} Cross-peak amplitude vs.\ $T$.  
}
\label{zeroquant}
\end{figure}

Figure \ref{zeroquant} shows an analysis of these effects in the 5-nm-barrier and 10-nm-barrier quantum well samples. Zero-quantum coherences were analyzed by slicing through the lower-left cross-peak (i.e., the cross-peak for which $|\hbar \omega_\tau| > |\hbar \omega_t|$). In both cases, cross-peak amplitude oscillations are clearly visible. These amplitude oscillations are due to interference between population terms and zero-quantum coherences. They beat at the difference frequency between the two excited states. These results between direct and indirect excitons serve as conclusive evidence for coherent interwell coupling between direct and indirect excitons, expanding upon the findings of Ref.~\cite{Nardin2014} where cross-peak oscillations were reported for interactions between direct excitons.

The zero-quantum decoherence rates are relatively fast compared to single-quantum coherence rates. For the 5-nm-barrier sample, we measure this decoherence rate to be $0.7 \pm 0.1$ ps$^{-1}$, and for the 10-nm-barrier sample, we measured to be $1.1 \pm 0.2$ ps$^{-1}$ [Decay rates were extracted by fitting the peak amplitude vs.\ $T$ curve to a decaying exponential of the form $f(T) = Ae^{-\gamma T}\cos(\omega T + \phi)+B$]. In contrast, the homogeneous single-quantum dephasing rates in the same data set are only $0.15 \pm 0.01$ ps$^{-1}$, and $0.13 \pm 0.01$  ps$^{-1}$ (for the 5-nm-barrier and 10-nm-barrier samples, respectively), as extracted by fitting diagonal single-quantum spectral peaks to an inhomogeneously broadened two-dimensional Lorentzian function \cite{Siemens2010}. These observations signify that transition energy fluctuations for the low-energy excitons in these samples are uncorrelated or anticorrelated. This is similar to the kind of dynamics that have been observed for excitons in GaAs/AlGaAs quantum wells \cite{Spivey2008} and CdTe/CdMgTe quantum wells \cite{Moody2014}, but very different from the dynamics observed for excitonic transitions in InAs quantum dot ensembles \cite{Moody2013b}. Anticorrelated energy level fluctuations can arise in principle from strain effects in semiconductors \cite{Lee1988}, and so the results may indicate the presence of electron-electron and electron-phonon interactions in the sample as a source of transient strain \cite{Spivey2008}.

\section{Discussion and Conclusions}
\label{sec:conclusions}

In summary, we have conducted a study of excitonic coupling and tunneling effects in InGaAs double quantum well samples exhibiting 5-nm, 10-nm, and 30-nm barriers between wells using a combination of multidimensional coherence spectroscopy (MDCS) and photoluminescence excitation spectroscopy (PLE) techniques. We have observed the emergence of optically accessible spatially indirect excitons in samples of reduced barrier thickness. Analyses of the exciton peak phases indicate that EID effects become increasingly dominant over EIS effects as exciton energies approach quantum well continuum states. This is true regardless of whether the excitons are direct or indirect and regardless of whether or not the quantum wells are coupled. Analyses of zero-quantum coherences in comparison to single-quantum coherences indicate that energy level fluctuations are either uncorrelated or anticorrelated.

The observations indicate that continuum states drive many-body interactions in ways that are different from the interactions between the excitons themselves, resulting in dephasing effects dominating over energy level shifts. From a theoretical standpoint, an EID-dominated effect for continuum-state interactions is reasonable to understand and perhaps even expected because of the large number of relevant interaction channels and because of the high degree of randomness involved. From an experimental standpoint, such findings are consistent with the findings of previous studies including an observation of enhanced dephasing effects for light-hole excitons overlapping with heavy-hole continuum states in GaAs quantum wells \cite{Honold1992}, an observation that EID dominates over EIS as the waiting time $T$ increases in MDCS measurements of GaAs quantum wells \cite{Turner2012}, and an observation of EID dominated many-body effects in InGaAs double quantum wells where narrow-well exciton energies overlap with  wide-well continuum states \cite{Nardin2014} (note that Ref.~\cite{Nardin2014} examined quantum wells of 8-nm and 10-nm thickness whereas those here are 9-nm and 10-nm thickness). The present study expands upon previous work by showing the universality of these properties in semiconductor quantum wells as relevant to direct and indirect excitons alike, and by illustrating that the impact of continuum states can be an exciton-energy dependent effect.

The results open the possibility of tunably controlling exciton properties in semiconductor quantum wells for desired functionality. For example, excitons could be engineered in InGaAs quantum wells of 9-nm and 10-nm thicknesses to interact across quantum wells with reduced decoherence properties. These properties in turn could be used to generate enhanced forms of coherent light sources. In broader terms, they may be relevant to improving light-emitting diodes and lasers where multiple quantum wells are grown in close proximity to enhance the optical emission efficiency, to developing quantum devices where it may be desirable to controllably tune the interactions between excitations using physical distance, and in understanding naturally occurring systems like photosynthetic complexes where exciton dynamics and interactions are of paramount importance in shaping light-matter interactions.

\appendix

\section{Many-body Effects as MDCS Lineshape Signatures}
\label{sec:phaseshifts}

The impact of many-body effects on resonance lineshapes in MDCS are discussed at length in many places, including Refs.\ \cite{Li}, \cite{Nardin2014}, and \cite{Li2006}. Here, we provide a summary of the treatments in these resources and a derivation of the connection between the effects and resonance peak phase shifts as discussed and analyzed in association with Figs.~\ref{phaseresolved} and \ref{phaseresolved2} of the main text.

\begin{figure}[tb]\centering\includegraphics[width=3.375in]{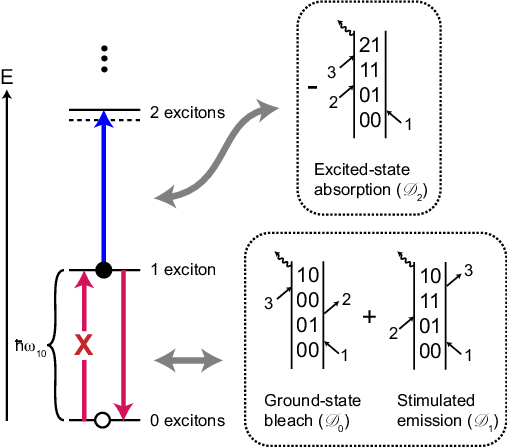}
\caption{Bosonic-ladder-style energy-level diagram for an excitonic excitation spectrum, and associated double-sided Feynman diagrams leading to the MDCS signal associated with this system when measured in the perturbative limit.}
\label{feynmandiagrams}
\end{figure}

Excitons are bosons, so it is appropriate in certain contexts to model their excitation spectrum as a ladder of nearly equally spaced energy levels \cite{Singh2016a} with each level corresponding to the excitonic occupation number as illustrated in Fig.~\ref{feynmandiagrams}. There are accordingly three different effects leading to the MDCS third-order rephasing spectrum signal output. Two of these signal contributions are positive (see the double-sided Feynman diagrams in Fig.~\ref{feynmandiagrams} labeled $\mathscr{D}_0$ and $\mathscr{D}_1$), and they originate from a combination of stimulated emission and ground state bleaching effects. One of the signal contributions is negative (see the diagram in Fig.~\ref{feynmandiagrams} labeled $\mathscr{D}_2$), and it originates from an excited-state absorption effect in which a singly-excited exciton state gets boosted into a doubly-excited exciton state. In the absence of many-body interactions, the positive and negative signal contributions very nearly cancel, but the cancellation becomes increasingly imperfect as many-body effects turn on. In fact, these many body effects are believed to dominate the signal in almost all semiconductor quantum well cases. Many-body effects can become manifest in one of two ways. The first of these is excitation-induced shift (EIS), which originates from a picture in which repulsive exciton-exciton interactions tend to drive the two-exciton state to an energy slightly higher than twice the one-exciton state. The overall result is a 90$^\circ$ phase-shifted MDCS signal contribution. Second is the possibility of excitation-induced dephasing (EID), which originates from a picture in which exciton-exciton interactions tend to cause the two-exciton state to dephase faster than the one-exciton state. In frequency-space this enhanced dephasing interaction leads to an asymmetry in resonance lineshape widths and a 0$^\circ$ phase-shifted MDCS signal contribution. 

%%%%%%%%%%%%%%%%%%%%%%%%%%%%%%%%%
%Fig8
%%%%%%%%%%%%%%%%%%%%%%%%%%%%%%%%%

Casting these concepts into equations based on density matrix perturbation theory \cite{Li}, the relevant MDCS signal is $S = \mathscr{D}_0 + \mathscr{D}_1 + \mathscr{D}_2$ or, because bleaching and stimulated signal contributions are equal,
\begin{align}
S &= 2\mathscr{D}_1 + \mathscr{D}_2
\end{align}
with
\begin{align}
\mathscr{D}_1 &= A|\mu|^4 \theta(\tau) e^{-i \Omega_{01} \tau} \theta(t) e^{-i \Omega_{10} t}
\intertext{and}
\mathscr{D}_2 &= -2A|\mu|^4 \theta(\tau) e^{-i \Omega_{01} \tau} \theta(t) e^{-i (\Omega_{10} + \Delta) t},
\end{align} 
where $A$ is a constant, $\mu$ is the light-matter interaction dipole moment, $\theta(x)$ is a Heaviside theta function, $\Omega_{10} = - \Omega_{01}^* = \omega_{10} - i\gamma $ is a complex resonance frequency associated with the singly excited exciton state energy $\hbar \omega_{10}$ and lifetime $1/\gamma$, $\tau$ is the time delay between first-order and second-order MDCS pulses, $t$ is the time following the third-order MDCS pulse, and where the parameter $\Delta = \textrm{EIS} - i\textrm{EID}$. Adding up terms and Fourier-transforming the result gives 
\begin{align}
S(\omega_t,\omega_\tau) &\propto \left( \frac{i}{\omega_\tau - \Omega_{01}} \right)\left( \frac{i}{\omega_t - \Omega_{10}} \right) \nonumber \\
&- \left( \frac{i}{\omega_\tau - \Omega_{01}} \right) \left( \frac{i}{\omega_t - [\Omega_{10}+\Delta]} \right).
\end{align}
If we assume that $|\Delta| \ll |\Omega_{10}|$, then we can write the final part of the second term of the expression above as
\begin{align}
i ( \omega_t - \Omega_{10} &- \Delta )^{-1} \nonumber \\
&= \left( \frac{i}{\omega_t - \Omega_{10}} \right) \left(1 - \frac{\Delta}{\omega_t - \Omega_{10}} \right)^{-1} \\
&\approx \left( \frac{i}{\omega_t - \Omega_{10}} \right) \left(1 + \frac{\Delta}{\omega_t - \Omega_{10}} \right).
\end{align}
Thus,
\begin{align}
S(\omega_t,\omega_\tau) &\approx A|\mu|^4 \left( \frac{i}{\omega_\tau - \Omega_{01}} \right)
\left( \frac{-i\Delta}{[\omega_t - \Omega_{10}]^2} \right). \label{sapprox}
\end{align}
Now, on resonance, all of the factors in Eq.~(\ref{sapprox}) except for $i\Delta$ evaluate to something purely real-valued and positive, so that phase of $S(\omega_t,\omega_\tau)$ at the center of the resonance is given by the phase of $i\Delta$. Writing this explicitly out in terms of the EIS and EID terms, we have
\begin{align}
i \Delta &= i (\textrm{EIS} - i\textrm{EID} ) \\
&= \textrm{EID} + i \textrm{EIS},
\end{align}
so that, on resonance, $S(\omega_t,\omega_\tau) \propto B e^{i\phi}$, with $B = \sqrt{ (\textrm{EID})^2+(\textrm{EIS})^2}$ and 
\begin{equation}
\phi = \tan^{-1}\left( \frac{\textrm{EIS}}{\textrm{EID}} \right).
\end{equation}
Pure EID thus leads to no phase shift at the center of resonance peaks, pure EIS leads to a 90$^\circ$ phase shift at the center of resonance peaks, and equal contributions of EID and EIS lead to a 45$^\circ$ phase shift at the center of resonance peaks. 

\begin{acknowledgments}
C.L.S. acknowledges partial support from an NRC Research Associateship award at NIST and support from NSF Grant No.\ 2003493.
\end{acknowledgments}

\bibliography{/Users/Chris/Documents/TexDocuments/clsbib2}
%\bibliography{clsbib2}

\end{document}